\documentclass{egpubl}
\usepackage{eurovis2022}

\SpecialIssuePaper         

\usepackage[T1]{fontenc}
\usepackage{dfadobe}

\biberVersion
\BibtexOrBiblatex
\usepackage[backend=biber,bibstyle=EG,citestyle=alphabetic,backref=true]{biblatex}
\AtEveryBibitem{ 
    \clearfield{urlyear}
    \clearfield{urlmonth}
    \clearfield{issn}
    \clearfield{isbn}
    \clearlist{location}
    \clearlist{language}
    \clearfield{pages}
    \clearfield{volume}
    \clearfield{number}
    \clearfield{series}
}
\electronicVersion
\PrintedOrElectronic
\ifpdf \usepackage[pdftex]{graphicx} \pdfcompresslevel=9
\else \usepackage[dvips]{graphicx} \fi

\usepackage{egweblnk}

\usepackage[capitalise, nameinlink]{cleveref}
\usepackage{array}
\usepackage{booktabs}
\usepackage{hhline}
\usepackage{multirow}
\newcolumntype{P}[1]{>{\centering\arraybackslash}p{#1}}
\usepackage{subcaption}

\usepackage{soul} 
\usepackage{dirtytalk} 
\usepackage{csquotes} 
\usepackage{censor} 
\usepackage[table]{xcolor}

\usepackage[labelfont=bf,textfont=it]{caption}
\usepackage[labelfont=bf,textfont=it]{subcaption}

\usepackage[nolist,nohyperlinks, smaller]{acronym}
\begin{acronym}
    \acro{SVG}{Scalable Vector Graphics}
    \acro{NL}{natural language}
    \acro{JSON}{JavaScript Object Notation}
    \acro{HTML}{HyperText Markup Language}
    \acro{AI}{Artificial Intelligence}
    \acro{CSS}{Cascading Style Sheets}
    \acro{HTML5}{HTML5}
    \acro{JSON}{JSON}
    \acro{HCI}{Human-Computer Interaction}
    \acro{ARIA}{Accessible Rich Internet Application}
    \acro{W3C}{World Wide Web Consortium}
    \acro{WCAG}{Web Content Accessibility Guidelines}
    \acro{WAI}{Web Accessibility Initiative}
    \acro{JAWS}{Job Access With Speech}
    \acro{NVDA}{NonVisual Desktop Access}
    \acro{CHI}{Conference on Human Factors in Computing Systems}
    \acro{BLV}{Blind Low-Vision}
    \acro{AT}{Assistive Technology}
    \acro{BANA}{Braille Authority of North America}
    \acro{APH}{American Printing House for the Blind}
    \acro{CS}{Computer Science}
    \acro{UD}{Universal Design}
    \acro{PD}{Participatory Design}
    \acro{ID}{Inclusive Design}
    \acro{PWD}{People with Disabilities}
    \acro{CV}{Computer Vision}
    \acro{NLP}{Natural Language Processing}
    \acro{NVDA}{NonVisual Desktop Access}
    \acro{STEM}{science, technology, engineering, and math}
    \acro{COVID19}{COVID-19}
    \acro{NLIs}{Natural Language Interfaces}
    \acro{SVT}{Sentence Verification Technique}
\end{acronym}

\definecolor{anti-flashwhite}{rgb}{0.95, 0.95, 0.96}
\definecolor{danielHajasQuote}{rgb}{0.9,1,0.96}
\definecolor{screenReaderQuote}{rgb}{1,1,0.8}
\definecolor{participantQuote}{rgb}{0.93,0.92,1}

\newcommand{\hlc}[2][yellow]{{\sethlcolor{#1}\hl{#2}}}
\newcommand{\hlquote}[2][black]{\textcolor{black}{\hlc[participantQuote]{\emph{``#2''}}}}
\newcommand{\dhquote}[2][black]{\textcolor{black}{\hlc[danielHajasQuote]{\emph{``#2''}}}}
\newcommand{\srquote}[2][black]{\textcolor{black}{\hlc[screenReaderQuote]{\emph{``#2''}}}}

\StopCensoring

\title{Rich Screen Reader Experiences for Accessible Data Visualization}

\renewcommand{\dagger}{*} 
\author[J. Zong, C. Lee, \& A. Lundgard et al.]
{\parbox{\textwidth}{\centering
\vspace{-20mm}
Jonathan Zong\thanks{Equal contribution}$^{1}$\orcid{0000-0003-4811-4624}, 
Crystal Lee\footnotemark[1]$^{1}$\orcid{0000-0001-6672-9118},
Alan Lundgard\footnotemark[1]$^{1}$\orcid{0000-0002-1352-3615},
JiWoong Jang$^{2}$\orcid{0000-0003-0469-9501},
Daniel Hajas$^{3}$\orcid{0000-0002-2811-1197},
Arvind Satyanarayan$^{1}$\orcid{0000-0001-5564-635X}
        }
        \\
{\parbox{\textwidth}{\centering
\vspace{-20mm}
$^1$~Massachusetts Institute of Technology~
$^2$~Carnegie Mellon University~
$^3$~University College London~
       }
}
}

\DeclareUnicodeCharacter{2212}{-}
\begin{document}


\maketitle

\begin{abstract}
    Current web accessibility guidelines ask visualization designers to support screen readers via basic non-visual alternatives like textual descriptions and access to raw data tables.
But charts do more than summarize data or reproduce tables; they afford interactive data exploration at varying levels of granularity\,---\,from fine-grained datum-by-datum reading to skimming and surfacing high-level trends.
In response to the lack of comparable non-visual affordances, we present a set of rich screen reader experiences for accessible data visualization and exploration.
Through an iterative co-design process, we identify three key design dimensions for expressive screen reader accessibility:
\emph{structure}, or how chart entities should be organized for a screen reader to traverse; 
\emph{navigation}, or the structural, spatial, and targeted operations a user might perform to step through the structure;
and, \emph{description}, or the semantic content, composition, and verbosity of the screen reader's narration.
We operationalize these dimensions to prototype screen-reader-accessible visualizations that cover a diverse range of chart types and combinations of our design dimensions.
We evaluate a subset of these prototypes in a mixed-methods study with 13 blind and low vision readers.
Our findings demonstrate that these designs help users conceptualize data spatially, selectively attend to data of interest at different levels of granularity, and experience control and agency over their data analysis process.
An accessible HTML version of this paper is available at:~\href{http://vis.csail.mit.edu/pubs/rich-screen-reader-vis-experiences}{http://vis.csail.mit.edu/pubs/rich-screen-reader-vis-experiences}.
\end{abstract}

\section{Introduction}\label{section:Introduction}

Despite decades of visualization research and recent legal requirements to make web-based content accessible~\cite{w3c_web_2018, higgins_supreme_2019}, web-based visualizations remain largely inaccessible to people with visual disabilities. 
Charts on mainstream publications are often completely invisible to screen readers (an assistive technology that transforms text and visual media into speech) or are rendered as incomprehensible strings of \emph{``graphic graphic graphic''}~\cite{sharif_understanding_2021, sarah_l_fossheim_how_2020}.
Current accessibility guidelines ask visualization designers to provide textual descriptions of their graphics via alt text (short for alternative text) and link to underlying data tables~\cite{gould_effective_2008, w3c_wai_2019}.
However, these recommendations do not provide modes of information-seeking comparable to what sighted readers enjoy with interactive visualizations. 
For instance, well-written alt text can provide a high-level takeaway of what the visualization shows, but it does not allow readers to drill down into the data to explore specific sections. 
While tables provide readers with the ability to hone in on specific data points, reading data line-by-line quickly becomes tedious and makes it difficult to identify overall trends.

Developing rich non-visual screen reader experiences for data visualizations poses several unique challenges. 
Although visuomotor interactions (like hovering, pointing, clicking, 
and dragging) have been core to visualization research~\cite{dimara_what_2020}, screen readers redefine what interaction is for visualization.
Rather than primarily \emph{manipulating} aspects of the visualization or its backing data pipeline~\cite{yi_toward_2007, heer_interactive_2012, dimara_what_2020}, screen readers make \emph{reading} a visualization an interactive operation as well\,---\,users must intentionally perform actions with their input devices in order to cognize visualized elements.
Moreover, as screen readers narrate elements one-at-a-time, they explicitly linearize reading a visualization.
As a result, in contrast to sighted readers who can choose to selectively attend to specific elements and have access to the entire visualization during the reading process, screen reader users are limited to the linear steps made available by the visualization author and must remember (or note down) prior output conveyed by the screen reader.  
Despite these modality differences, studies have found that screen reader users share the same information-seeking goals as sighted readers: an initial holistic overview followed by comparing data points~\cite{sharif_understanding_2021}, akin to the information-seeking mantra of ``overview first, zoom and filter, and details on demand''~\cite{shneiderman_eyes_2003}.

In this paper, we begin to bridge this divide by conducting an iterative co-design process (co-author Hajas
is a blind researcher with relevant experience) prototyping rich and usable screen reader experiences for web-based visualizations.
We identify three design dimensions for enabling an expressive space of experiences: \emph{structure}, or how the different elements of a chart should be organized for a screen reader to traverse; \emph{navigation}, which describes the operations a user may perform to move through this structure; and, \emph{description}, which specifies the semantic content, composition, and verbosity of text conveyed at each step. 
We demonstrate how to operationalize these design dimensions through diverse accessible reading experiences across a variety of chart types.

To evaluate our contribution, we conduct an exploratory mixed-methods study with a subset of our prototypes and 13 blind or low vision screen reader users. 
We identify specific features that make visualizations more useful for screen reader users (e.g., hierarchical and segmented approaches to presenting data, cursors and roadmaps for spatial navigation) and identify behavior patterns that screen reader users follow as they read a visualization (e.g., constant hypothesis testing and validating their mental models). 

\section{Background and Related Work}
\label{section:RelatedWork}

\textbf{Screen Reader Assistive Technology.}
A screen reader is an assistive technology that conveys digital text or images as synthesized speech or braille output. 
Screen readers are available as standalone third-party software or can be built-in features of desktop and mobile operating systems.
A screen reader allows a user to navigate content linearly with input methods native to a given platform (e.g., touch on smartphones, mouse/keyboard input on desktop).
Content authors must generate and attach alt text to their visual content like images or charts in order for them to be accessible to screen reader users. 
Functionality and user experience differs across platforms and screen readers.
In this paper, however, we focus on interacting with web-based visualizations with the most widely used desktop screen readers (JAWS/NVDA for Windows, VoiceOver for Mac).

\textbf{Web Accessibility Standards.}
In 2014, the World Wide Web Consortium (W3C) adopted the Web Accessibility Initiative's Accessible Rich Internet Applications protocol (WAI-ARIA) which introduced a range of semantically-meaningful HTML attributes to allow screen readers to better parse HTML elements~\cite{mdn_contributors_aria_2021}.
In particular, these attributes allow a screen reader to convey the state of dynamic widgets (e.g., autocomplete is available for text entry), alert users to live content updates, and identify common sections of a web page for rapid navigation (e.g., banners or the main content). 
In 2018, the W3C published the WAI-ARIA Graphics Module~\cite{w3c_wai-aria_2018} with additional attributes to support marking up structured graphics such as charts, maps, and diagrams.
These attributes allow designers to annotate individual and groups of graphical elements as well as surface data values and labels for a screen reader to read aloud. 

\textbf{Accessible Visualization Design.}
In a recent survey, Kim et al.~\cite{kim_accessible_2021} describe the rich body of work that has explored multi-sensory approaches to visualization for multiple disabilities~\cite{goos_haptic_2001,hasty_guidelines_2011,kaper_data_1999,barrass_using_1999,wu_understanding_2021, lundgard_sociotechnical_2019}.
Here, we focus on screen reader output native to web-based interfaces for blind users (namely via speech).
Sharif et al.~\cite{sharif_understanding_2021} find that many web-based charts are intentionally designed to cause screen readers to skip over them.
For charts that a screen reader does detect, blind or low vision users nevertheless experience significant difficulties: these users spend 211\% more time interacting with the charts and are 61\% less accurate in extracting information compared to non-screen-reader users~\cite{sharif_understanding_2021}.
Despite the availability of ARIA, alt text and data tables remain the most commonly used and recommended methods for making web-based charts accessible to screen readers~\cite{gould_effective_2008, w3c_wai_2019, choi_visualizing_2019}. 
However, each of these three approaches comes with its own limitations.
Static alt text requires blind readers to accept the author's interpretation of the data; by not affording exploratory and interactive modes, alt text robs readers of the necessary time and space to interpret the numbers for themselves~\cite{lundgard_accessible_2021}.
Recent research also suggests that blind people have nuanced preferences for the kinds of visual semantic content conveyed via text~\cite{potluri_examining_2021, lundgard_accessible_2021}, and desire more interactive and exploratory representations of pictorial images~\cite{morris_rich_2018}.
Data tables, on the other hand, undo the benefits of abstraction that visualizations enable\,---\,they force readers to step sequentially through data values making it difficult to identify larger-scale patterns or trends, and do not leverage the structure inherent to web-based grammars of graphics~\cite{bostock_d3_2011, satyanarayan_vega-lite_2017}.
Finally, ARIA labels are not a panacea; even when they are used judiciously\,---\,a non-trivial task which often results in careless designs that cause screen readers to simply read out long sequences of numbers without any other identifiable information~\cite{sarah_l_fossheim_how_2020}\,---\,they present a fairly low expressive ceiling.
The current ARIA specification does not afford rich and nuanced information-seeking opportunities equivalent to those available to sighted readers.

There has been some promising progress for improving support for accessibility within visualization toolkits, and vice-versa for improving native support for charts in screen reader technologies. 
For instance, Vega-Lite~\cite{satyanarayan_vega-lite_2017} and Highcharts~\cite{highcharts_accessibility_2021} are beginning to provide ARIA support out-of-the-box.
Apple's VoiceOver Data Comprehension feature~\cite{davert_whats_2019} affords more granular screen reader navigation within the chart, beyond textual summaries and data tables, via four categories of selectable interactions for charts appearing in Apple's Stocks or Health apps. 
These interactions include \emph{Describe Chart}, which describes properties of the chart's construction, such as its encodings, axis labels, and ranges; \emph{Summarize Numerical Data}, which reports min and max data values, and summary statistics like mean and standard deviation; \emph{Describe Data Series}, which reports the rate-of-change/growth of a curve, trends, and outliers; and \emph{Play Audiograph}, which plays a tonal representation of the graph's ascending/descending trend over time~\cite{davert_whats_2019}.
While Apple's features are presently limited to single-line charts, SAS' Graphics Accelerator~\cite{sas_graphics_accelerator_sas_2018} supports a similar featureset (including sonification, textual descriptions, and data tables) but for a broader range of statistical charts including bar charts, box plots, contour plots, and scatter plot matrices.
Our work follows in the spirit of these tools but focuses on web-based visualizations rather than standalone- or platform-integrated software.
We go beyond what ARIA supports today to enable high-level and fine-grained screen reader interactions, and hope that our work will help inform ongoing discussions on improving web accessibility standards (e.g., via an Accessibility Object Model~\cite{boxhall_accessibility_2022}).

\section{Design Dimensions for Rich Screen Reader Experiences}
\label{section:DesignDimensions}

Currently, the most common ways of making a visualization accessible to screen readers include adding a single high-level textual description (via alt text), providing access to low-level data via a table, or tagging visualization elements with ARIA labels to allow screen readers to step through them linearly (e.g., as with Highcharts~\cite{highcharts_accessibility_2021}).
While promising, these approaches do not afford rich information-seeking behaviors akin to what sighted readers enjoy with interactive visualizations. 
To support systematic thinking about accessible visualization design, we introduce three design dimensions that support rich, accessible reading experiences: \textit{structure}, or how elements of the visualization should be organized for a screen reader to traverse; \textit{navigation}, or the mechanisms by which a screen reader user can move from one element to another; and \textit{description}, or what semantic content the screen reader conveys.

\textbf{Methods.} 
We began by studying the development of multi-sensory graphical systems, covering work in critical cartography~\cite{wiedel_tactual_1969,koch_state_2012}, blind education~\cite{aldrich_tactile_2001,godfrey_advice_2015}, tactile graphics~\cite{miesenberger_universal_2018,hasty_guidelines_2011,de_greef_interdependent_2021,amick_guidelines_1997,butler_technology_2021}, and multi-sensory visualization~\cite{martinez_accessible_2019,chundury_towards_2021,brock_usage_2010,baker_tactile_2016}. 
Drawing on conventions and literature on crip, reflective, and participatory design~\cite{hamraie_designing_2013,sengers_reflective_2005,costanza-chock_design_2020}, all authors began an iterative co-design process with \censor{Hajas}, who is a blind researcher with relevant expertise.
\censor{Hajas} is a screen reader user with a PhD in HCI and accessible science communication, but he is not an expert in visualization research. 
Co-design\,---\,particularly as encapsulated in the disability activism slogan, \emph{``Nothing about us, without us''}~\cite{costanza-chock_design_2020}\,---\,is important because it can eliminate prototypes that replicate existing tools, solve imaginary problems (i.e., by creating disability dongles~\cite{jackson_disability_2019}) or unintentionally produce harmful technology~\cite{shew_ableism_2020}.
To balance engaging disabled users while acknowledging academia's traditionally extractive relationship with marginalized populations~\cite{cornwall_what_1995}, we intentionally acknowledge \censor{Hajas} as both co-designer and co-author.
We believe that the distinction between co-designer\,---\,a phrase that often discounts lived experience as insufficiently academic\,---\,and researcher is minimal; technical, qualitative, and experiential expertise are all important components of this research. 
\censor{Hajas}' profile is a perfect example of the intersection between lived experience of existing challenges and solutions, academic experience of research procedures, and an interest in the science of visualization. 
While he does not represent all screen reader users, his academic expertise and lived experience uniquely qualify him to be both researcher and co-designer. 
Nevertheless, to incorporate a diverse range of perspectives, we recruited additional participants as part of an evaluative study (\S~\ref{section:Evaluation}).

Our work unfolded over 6 months and yielded 15 prototypes.
All authors met weekly for hour-long video conferences.
In each session, we would discuss the structure and affordances of the prototypes, often by observing and recording \censor{Hajas}' screen as he worked through them. 
We would also use these meetings to reflect on how the prototypes have evolved, compare their similarities and differences, and whiteboard potential design dimensions to capture these insights.
Following these meetings, \censor{Hajas} wrote memos detailing the motivations for each prototype, tagging its most salient features, summarizing the types of interactions that were available, enumerating questions that the prototype raises, and finally providing high-level feedback about its usefulness and usability.
In the following section, we liberally quote these memos to provide evidence and additional context for our design dimensions.

\subsection{Structure} 
We define \emph{structure} to mean an underlying representation of a visualization that organizes its data and visual elements into a format that can be traversed by a screen reader.
Through our co-design process, we identified two components important to analyzing accessible structures: their \textit{form}, or the shape they organize information into; and \textit{entities}, or which parts of the visualization specification are used to translate a chart into a non-visual structure.
Design decisions about form and entities are guided by considerations of \emph{information granularity}, or how many levels comprise the range between a high-level overview and individual data values.

\begin{figure*}[htb]
  \centering
  \includegraphics[width=\linewidth]{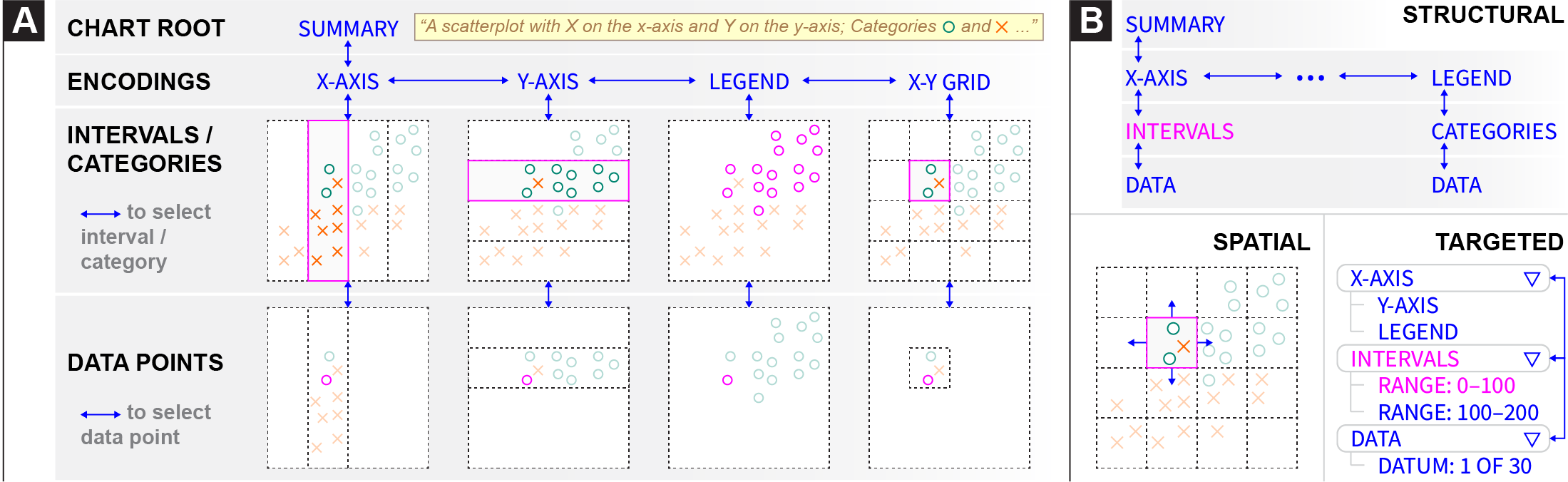}
  \caption{\label{fig:encoding}
          (a) An accessible visualization structure in the form of a tree and comprised of encoding entities.
          Solid \textcolor{magenta}{magenta} outlines indicate the location of the screen reader cursor.
          Solid \textcolor{blue}{blue} arrows between labels indicate available next steps via keyboard navigability (up, down, left, right).
          (b)
          Three ways of navigating accessible visualization structures: structural, spatial, and targeted.
          }
\vspace{-5mm}
\end{figure*}

\textbf{Form.} 
Accessible structures organize information about the visualization into different forms, including lists, tables, and trees.
Consider existing best practices and common approaches.
A rasterized chart with alt text is represented to a screen reader as a single node.
SVG-based visualizations can additionally be tagged with ARIA labels to describe the axes, legends, and individual data points.
Despite SVG's nesting, screen readers linearize these ARIA labels into a list structure so that the user can step through them sequentially.
Data tables, on the other hand, provide a grid structure for screen readers to traverse.
At each cell of the grid, the screen reader reads out a different textual description, allowing the user to explore a space by traversing the grid spatially (up, down, left, and right) instead of merely linearly.
Accessible visualization research has begun to explore the use of tree structures for storing chart metadata~\cite{weninger_asvg_2015}, but they remain relatively rare in practice.
Our prototypes primarily use trees as their branching and hierarchical organization allows users to browse different components of a visualization and traverse them at different levels of detail.

\textbf{Entities.} 
Where form refers to how nodes in a structure are arranged, entities instead refers to what aspects of the visualization the nodes represent.
These aspects can include:

\begin{itemize}
    \item \textit{Data}, where nodes in the structure represent individual data values or different slices of the data cube (e.g., by field, bins, categories, or interval ranges).
For example, in a data table, every node (i.e. cell) represents a data value designated by the row and column coordinates.
Depending on the form, data entities can be presented at different levels of detail.
For example, one prototype we explored represents a line chart as a binary tree structure (Fig.~\ref{fig:example-gallery}e): the root node represents the entire x-axis domain, and each left and right child node recursively splits the domain in half.
Users can traverse the tree downward to binary search for specific values or understand the data distribution.

    \item \textit{Encodings,} where nodes in the structure correspond to visual channels (e.g., position, color, size) that data fields map to.
For instance, consider Figure~\ref{fig:encoding}a which depicts the encoding structure of a Vega-Lite scatterplot. 
The visualization is specified as mappings from data fields to three visual encoding channels: \texttt{x}, \texttt{y}, and \texttt{color}.
Thus, the encoding structure, which here takes the form of a tree, comprises a root node that represents the entire visualization and then branches for each encoding channel as well as the data rectangle (x-y grid).
Descending into these branches yields nodes that select different categories or interval regions, determined by the visual affordances of the channel. 
For instance, descending into axis branches yields nodes for each interval between major ticks; x-y grid nodes represent cells in the data rectangle as determined by intersections of the axes gridlines; and legend nodes reflect the categories or intervals of the encoding channel (i.e., for nominal or quantitative data respectively). 
Finally, the leaves of these branches represent individual data values that fall within the selected interval or category. 
    \item \textit{Annotations,}
    where nodes in the structure represent the rhetorical devices a visualization author may use to to shape a visual narrative or guide reader interpretation of data (e.g., by drawing attention to specific data points or visual regions).
    Surfacing annotations in the visualization structure allows screen reader users to also benefit from and be guided by the author's narrative intent.
For example, Figure~\ref{fig:example-gallery}d illustrates an annotation tree structure derived from an example line chart with two annotations highlighting intervals in the temporal x-axis. The root of the tree has two children representing the two annotated regions.
The these two annotation nodes have a child node for each data point that is highlighted within the region of interest. 
\end{itemize}

\textbf{Considerations: Information Granularity.}
When might users prefer nested structures (i.e. trees) over flat structures (i.e., lists and tables)?
Like sighted users, screen reader users seek information by looking for an overview before identifying subsets to view in more detail~\cite{sharif_understanding_2021}.
Trees allow users to read summary information at the top of the structure, and traverse deeper into branches to acquire details-on-demand.
Kim et al. use the term \textit{information granularity} to refer to the different levels of detail at which an accessible visualization might reveal information~\cite{kim_accessible_2021}. 
They organize granularity into three levels: existence, overview, and detail.
\textit{Existence} includes information that a chart is present, but no information about underlying data.
\textit{Overview} includes summary information about data\,---\,e.g. axes, legends, and summary statistics like min, max, or mean\,---\,but not individual data points.
\textit{Detail} includes information about precise data values.

We use the root node to signal the existence of the tree, and deeper nodes in the tree reflect finer levels of granularity.
Branch nodes give an overview summary about the data underneath, providing information scent~\cite{pirolli_information_1999}, while leaf nodes map to individual data points.
In his feedback about the prototype shown in Figure~\ref{fig:encoding}, \censor{Hajas} wrote \dhquote{considering how difficult reading a scatterplot with a screen reader is due to its sequential reading nature, the tree structure makes the huge number of data points fairly readable}.

Entities are not mutually exclusive, and a structure might opt to surface different entities in parallel branches. 
We prototyped a version of Figure~\ref{fig:example-gallery}d which placed an encoding tree and annotation tree as sibling branches under the root node.
Users could descend down a given branch, and switch to the equivalent location in the other branch at will.
These design decisions are motivated by findings in prior work: by placing encodings and annotations as co-equal branches, we produce a structure that preserves the agency of screen reader users either to start with the narrative arc of annotations, or follow it after having the chance to interpret the data for themselves~\cite{lundgard_accessible_2021}.
As \censor{Hajas} confirms \dhquote{Depending on my task, either the encoding or annotation tree could be more important. If my task involved checking population growth in the last 100 years, I would start with the encodings. If I were to look for sudden changes in population numbers, such war-time mortality effects, I would start exploring the annotations, then tunnel back to the other tree.}

\subsection{Navigation}
\label{section:Navigation}

Screen reader users need ways to traverse accessible structures to explore data or locate specific points.
When browsing a webpage, screen readers provide a cursor that represents the current location in the page.
Users use keyboard commands to step the cursor backward and forward in a sequential list of selectable items on the page, or jump to important locations such as headers and links.
Through our prototyping process, we developed three ways of navigating through an accessible structure: \textit{structural navigation}, \textit{spatial navigation}, and \textit{targeted navigation} (Fig.~\ref{fig:encoding}b).
A key concern across these navigation schemes is reducing a user's cognitive load by affording a sense of the \emph{boundaries} of the structure.

\textbf{Structural Navigation.}
Structural navigation refers to ways users move within the accessible structure.
We identify two types of structural navigation.
\textit{Local navigation} refers to step-by-step movements between adjacent nodes in the structure. This includes moving up and down levels of a hierarchy, or moving side to side between sibling elements.
\textit{Lateral navigation} refers to movement between equivalent nodes in adjacent sub-structures.
For example, Fig.~\ref{fig:example-gallery}a depicts a multi-view visualization with six facets.
When the cursor is on a Y-axis interval for the first facet, directly moving to the same Y-axis interval on the second facet is a lateral move. 

\textbf{Spatial Navigation.}
Sometimes users want to traverse the visualization according to directions in the screen coordinate system. 
We refer to this as spatial navigation.
For example, when traversing part of an encoding structure that represents the visualization's X-Y grid, a downward structural navigation would go down a level into the currently selected cell of the grid, showing the data points inside the cell. 
A downward spatial navigation, in contrast, would move to the  grid  cell below the current one\,---\,i.e. towards the bottom of the Y-axis. 
Spatial navigation is also useful when navigating lists of data points, which may not be sorted by X or Y value in the encoding structure. 
Where a leftward structural navigation would move to the previous data point in the structure, a leftward spatial navigation would move to the point with the next lowest X value.


\textbf{Targeted Navigation.}
Navigating structurally and spatially requires a user to maintain a mental map of where their cursor is relative to where they want to go. 
If the user has a specific target location in mind, maintaining this mental map in order to find the correct path in the structure to their target can create unnecessary cognitive load. 
We use targeted navigation to refer to methods that only require the user to specify a target location, without needing to specify a path to get there. 
For example, the user might open a list of locations in the structure and select one to jump directly there. 
Screen readers including JAWS and VoiceOver implement an analogous form of navigation within webpages.
Instead of manually stepping through the page to find a specific piece of content, users can open a menu with a list of locations in the page.
These locations are defined in HTML using ARIA landmark roles, which can designate parts of the DOM as distinct sections when read by a screen reader.
When a screen reader user open the list of landmarks and selects a landmark, their cursor moves directly to that element.

\textbf{Considerations: Boundaries \& Cognitive Load.}
 Screen reader users only read part of the visualization at a time, akin to a sighted user reading a map through a small tube \cite{hasty_guidelines_2011}.
 How do they keep track of where they are?
 In our co-design process, we found it easiest for a user to remember their location relative to a known starting point, which is corroborated by literature on developing spatial awareness for blind people \cite{wiedel_tactual_1969,li_editing_2019,chundury_towards_2021}. \censor{Hajas} noted the prevalence of the \texttt{Home} and \texttt{End} shortcuts across applications for returning to a known position in a bounded space (e.g. the start/end of a line in a text editor).
 We also found that grouping data by category or interval was helpful for maintaining position. \censor{Hajas} noted that exploring data within a bounded region was like entering a room in a house.
 In his analogy, a house with many smaller rooms with doors is better than a house with one big room and no doors. 
 Bounded spaces alleviate cognitive load by allowing a user to maintain their position relative to entry points.
 
 Comparing navigation techniques, \censor{Hajas} noted that spatial felt \dhquote{shallow but broad} while targeted felt \dhquote{deep but narrow.}
 While he expressed a personal preference for deep-narrow structures, he nevertheless \dhquote{would not give up [spatial navigation] because it makes me believe I'm actually interacting with a visualization.}
 This insight demonstrates the value of offering multiple complementary navigation techniques.
 Moreover, while targeted navigation facilitates quick searching and doesn't require the user to maintain a mental map to find specific data points, structural and spatial exploration enable more open-ended data exploration. It also provides a mechanism for establishing common ground with sighted readers (e.g., allowing both blind and sighted readers to understand a line segment as being ``above'' or ``higher'' than another).
 
\subsection{Description}
When a user navigates to a node in a structure, the screen reader narrates a description associated with that node.
For example, when navigating to the chart's legend, the screen reader output might articulate visual properties of the chart's encoding: \srquote{Category O has color encoding green; X has color encoding orange} (Figure~\ref{fig:encoding}). 
Or, if that visual semantic content isn't relevant to understanding the data, it might ignore the color: \srquote{each datum belongs to either Category O or X.}
The \textit{content}, \textit{composition}, and \textit{verbosity} of the description can affect a user's comprehension of the data.
Designers must consider \textit{context \& customization} when describing charts.


\textbf{Content.}
Semantic \emph{content} is the meaningful information conveyed not only through natural language utterances, but also through the visualization (a graphical language~\cite{bertin_semiology_1983}).
Because graphics convey myriad different kinds of content, the challenge of natural language description is to convey information that is not only commensurate with what the chart expresses via graphical language, but also useful to its readers.
Accessible chart description guidelines from WGBH~\cite{gould_effective_2008}, W3C~\cite{w3c_wai_2019}, and others~\cite{jung_communicating_2021} offer prescriptions for conveying specific content for blind readers (such as the chart's title, axis encodings, and noteworthy trends).
Lundgard and Satyanarayan expand the scope of these guidelines with a more general conceptual model of four levels of semantic content: \emph{chart construction properties} (e.g., axes, encodings, marks, title); \emph{statistical concepts and relations} (e.g., outliers, correlations, descriptive statistics); \emph{perceptual and cognitive phenomena} (e.g., complex trends, patterns); and \emph{domain-specific insights} (e.g., socio-political context relevant to the data)~\cite{lundgard_accessible_2021}.

Decoupling a chart's semantic content from its visual representation helps us better understand what data representations afford for different readers.
For instance, Lundgard and Satyanarayan find that what blind readers report as most useful in a chart description is not a straightforward translation of the visual data representation.
Specifically, simply listing the chart's encodings is much less useful to blind readers than conveying summary statistics and overall trends in the data~\cite{lundgard_accessible_2021}.
As \censor{Hajas} noted, \dhquote{I want to see the global trend, which is why sighted people rely on visualization.} For instance, for a stock market chart the reader \dhquote{might see the overview from first to last data points, and then zoom into an outlier in the middle.}
These findings suggest opportunities interleave different kinds of content at different levels of a hierarchical structure to yield richer, more useful screen reader navigation. 
For example, injecting summary statistics (say, the existence of outliers within a particular subcategory of the data) higher up in the chart's tree structure (e.g., at the \texttt{legend} encoding node) might afford ``scent'' for ``information foraging''~\cite{pirolli_information_1999}, or further exploration down a particular branch (data subcategory) of the tree.
Or, if navigating in a targeted fashion, the user might be afforded the option to directly navigate to outliers without traversing the tree.

\textbf{Composition.}
The usefulness of a description depends not only on the content conveyed by its constituent sentences, but also on its \emph{composition}: how those sentences are ordered in relation to each other.
For example, during our co-design process, \censor{Hajas} found that when navigating a chart's tree structure, the screen reader output could quickly become redundant, affecting how quickly and efficiently he could pick out the meaningful information at each node.
For instance, the utterance \srquote{Category: O, Point 3 of 15, x = 5, y = 12} and the utterance \srquote{x = 5, y = 12, Category: O, Point 3 of 15} afford significantly different experiences for a user who wishes to quickly scan through individual data points.
In the first utterance, the reader immediately receives content that helps to situate them in a broader data context, namely data labeled as \say{Category: O} at the \texttt{legend} node.
In the second utterance, the reader immediately receives datum-specific content that helps to rapidly explore the fine-grained details within that data context.
Whether a reader prefers one compositional ordering to another will depend on the task they are attempting to accomplish.
As \censor{Hajas} noted \dhquote{I like the label at the beginning of the information, saying at which level of the tree I am at. It is important for knowing where I am. It is also great that this information is only spoken out when I change level, but not when I navigate laterally.}
These compositional choices are highly consequential for readers' experience, when they must repeatedly read nearly-identical utterances while navigating a structure.

\textbf{Verbosity.}
Whereas composition refers to the {ordering} of content, \emph{verbosity} refers to {how much} content the screen reader conveys.
More content is not always better.
As \censor{Hajas} noted of Apple's Data Comprehension feature~\cite{davert_whats_2019}: \dhquote{It can sometimes be too much information all at once, if it starts reading out all of the data. This is very difficult if you're interested in some data points that are in the middle. It is very play-or-stop.}
Depending on the screen reader software, a user may be afforded control over how much content is conveyed.
For instance, JAWS offers high, medium, and low verbosity levels~\cite{freedom_scientific_jaws_2021}.
At higher verbosity the screen reader announces more structural, wayfinding content (e.g. the start and end of regions).
For data tables, verbosity configurations can affect whether the table size is read as part of the description, and whether row and column labels are repeated for every cell.
Descriptions of nodes in an encoding structure might analogously include information about the path from the root\,---\,for example, by reminding the user that they are reading Y-axis intervals.
These repetitions can help users remember their location within a structure, but additional verbosity is less efficient for comprehending the data quickly.

\textbf{Considerations: Context \& Customization.}
Apart from its constituent parts (content, composition, verbosity), a description's usefulness also depends on the \emph{context} in which it is read: namely, the reader's task or intent, and familiarity with the data interface.
The same description might be useful in some situations, but relatively useless in others.
A reader's information needs are fundamentally context-sensitive.
For example, as \censor{Hajas} noted, when reading a news article, it may be satisfactory to accept a journalist's description of the data on good faith. But, when reviewing scientific research, \dhquote{I don't necessarily want to just believe what is said in the text, I want to check and double-check the authors' claims. Go down to the smallest numbers in the analysis. I want to be able to look at the confusion matrix and see if they made a mistake or not.}
This targeted verification requires a description to afford users with precise look-up capabilities, in contrast to descriptions that may be generated when browsing or exploring the data.

This context-sensitivity reveals an important aspect of usability: a user's familiarity (or lack thereof) with the data interface.
Wayfinding content (e.g., \srquote{Legend. Category: O.}) can help a user remember their location in a structure, and may be useful while they assemble a mental map of the visualization.
But, as they become accustomed to the interface and visualization, such descriptions may prove cumbersome.
Because user needs depend on their task, preferences, and familiarity, interfaces might afford personalization and customization to facilitate context-sensitive description.

\section{Example Gallery}
\label{section:ExampleGallery}

\begin{figure*}[!htb]
  \centering
  \includegraphics[width=\linewidth]{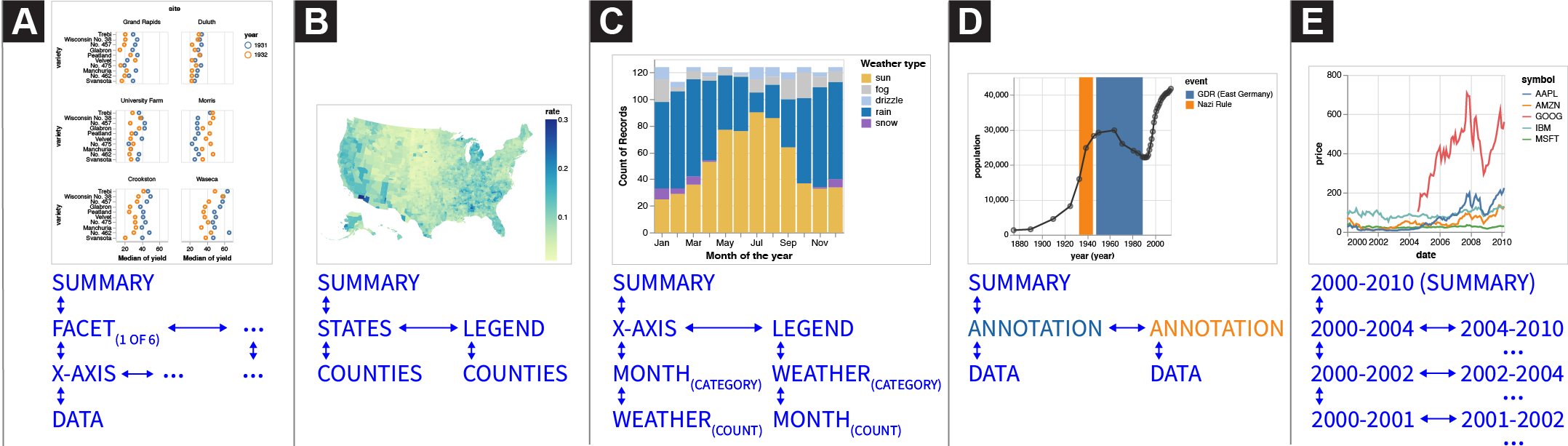}
  \vspace{-5mm}
  \caption{Example structural and navigational schemes generated as part of our co-design process, and applied to diverse chart types.}
  \label{fig:example-gallery}
\vspace{-5mm}
\end{figure*}

Our co-design process yielded prototypes that demonstrate a breadth of ways to operationalize our design dimensions.
Figure~\ref{fig:example-gallery} excerpts some of our highest-fidelity prototypes, implemented on top of Vega-Lite~\cite{satyanarayan_vega-lite_2017}.
As deeply nested structures and dynamic content are not well-supported by ARIA, we implemented our designs as in-memory data structures.
Event listeners update the user's position in the structure on keypress, and write text descriptions to an ARIA-Live region (an ARIA role typically used for temporary notifications).
To establish common ground with sighted users, we also render the visualization graphically.
The user's position in the tree drives a Vega-Lite selection that highlights points when the screen reader user is attending to them.

For every prototype, the \texttt{up}, \texttt{down}, \texttt{left}, and \texttt{right} arrow keys enable structural navigation (moving up or down a level, or stepping through siblings respectively). 
For example, within the facet level of Fig.~\ref{fig:example-gallery}(a), the user can press \texttt{left} or \texttt{right} keys to move between the six subplots of the multiview chart.
On charts that contain a node representing the x-y grid, users can also use the \texttt{WASD} keys to spatially navigate the grid and data points within that branch (mimicking an interaction found in video games). 

These prototypes highlight different compositions of structures and navigation schemes. 
Fig.~\ref{fig:example-gallery}(a) includes \texttt{shift+left} and \texttt{shift+right} for lateral navigation across facets: pressing these keys at any node within a facet branch will navigate to the same location under an adjacent branch (subplot).
With the chloropleth (Fig.~\ref{fig:example-gallery}(b)), we group data in the encoding structure by U.S. state; users can then drill down into counties across either this branch or the legend one.
Fig.~\ref{fig:example-gallery}(c) offers two different paths for drilling down: month first, or weather first.
Fig.~\ref{fig:example-gallery}(d) structures the tree by annotations rather than encoding: users can descend into the time intervals designated by the orange and blue rectangles, and view points within those intervals.
Finally, Fig.~\ref{fig:example-gallery}(e) organizes its tree in terms of data, offering a binary search structure through the years.

\section{Evaluation}
\label{section:Evaluation}

To evaluate our contribution, we conducted 90-minute Zoom studies with 13 blind and low vision participants. 
Participants were asked to explore three prototype accessible screen reader experiences, shown one after another each with a different dataset. 
The goal of our evaluation was not 
to determine which particular combination of design elements was ``best,'' but rather to be exploratory\,---\,to compare the relative strengths and advantages of instantiations of our design dimensions, and understand how they afford different modes of accessible data exploration.

\subsection{Evaluation Setup \& Design}

Following Frøkjær and Hornbæk's Cooperative Usability Testing (CUT) method~\cite{frokjaer_cooperative_2005}, \censor{Zong} and \censor{Lee} conducted each session by alternating between the role of guide (i.e., talking to the user and explaining the prototype) and logger (i.e., keeping track of potential usability problems, interpreting the data to prepare for becoming the guide).
We began each session with a semi-structured interview to understand participants' current experiences with data and the methods they use to make inaccessible forms of data representation usable (script included in supplementary material). 
The rest of the session focused on each of the three prototypes in turn, with each condition split into two phases: interaction and interpretation.
In the interaction phase, \censor{Zong or Lee} guided participants through the prototypes and asked participants to use them and comment on their process, in the style of Hutchinson et al.'s technology probes \cite{hutchinson_technology_2003}. 
Then, the authors switched roles and began a cooperative interpretation phase, where the authors and participants engaged in a constructive dialogue to jointly interpret the usability problems and brainstorm possible alternatives to the current prototype. 
In this method, participants influence data interpretation, allowing for more rapid analysis than traditional think-aloud studies as some analysis is built into each evaluation session with instant feedback or explanation from participants~\cite{frokjaer_cooperative_2005}.

\textbf{Prototypes.}
The in-depth nature of our cooperative interpretation sessions required us to balance the total number of prototypes evaluated (so that participants would have time to thoroughly learn and interact with each one) with a time duration appropriate for a Zoom session (limited to 90 minutes to avoid exhausting participants).
Accordingly, we selected the following three prototypes, each representing a different aspect of our design dimensions:

\newcommand{\Table}{\textsc{table}}
\newcommand{\Multiview}{\textsc{multi-view}}
\newcommand{\Target}{\textsc{target}}
\begin{itemize}
    \item \Table: An accessible HTML data table with all rows and three columns from the classic Cars dataset, in order to compare our prototypes with existing accessibility best practice.
    \item \Multiview: Becker's barley yield trellis display~\cite{becker_visual_1996} as shown in Fig.~\ref{fig:example-gallery}a. This prototype features local and lateral structural navigation via the arrow keys and with the shift modifier respectively, as well as spatial navigation via \texttt{WASD}.
    \item \Target: A single-view scatterplot, illustrated in Fig.~\ref{fig:encoding}, depicting the Palmer Penguins dataset~\cite{horst_palmerpenguins_2020}. In addition to structural and spatial navigation, targeted navigation is available via three dropdown menus corresponding to the structural levels.
\end{itemize}

\Table{} is our control condition, as it follows existing best practice for making data accessible to screen readers. 
\Multiview{} enables us to study how users move between levels of detail, and whether they could navigate and compare small multiple charts.
Finally, \Target{} allows us to compare how and when our participants use the three different styles of navigation (structural, spatial, and targeted).
We presented the prototypes in this sequence to all participants to introduce new features incrementally.

\textbf{Participants.}
We recruited 13 blind and low vision participants through our collaborators in the blind community and through a public call on Twitter.
Each participant received \$50 for a 90-minute Zoom session. 
We provide aggregate participant data following ethnographic practice to protect privacy and not reduce participants to their demographics~\cite{saunders_anonymising_2015}. 
Half of our participants were totally blind (n=7), while others were almost totally blind with some light perception (n=4) or low vision (n=2). 
Half of them have been blind since birth (n=7). 
Participants were split evenly between Windows/Chrome (n=7) and Mac/Safari (n=6). 
Windows users were also split evenly between the two major screen readers (JAWS, n=3; NVDA, n=4), while all Mac participants used Apple VoiceOver. 
These figures are consistent with recent surveys conducted by WebAIM which indicate that JAWS, NVDA, and VoiceOver are the three most commonly used screen readers~\cite{webaim_screen_2021}. Demographically, 70\% of our participants use he/him pronouns (n=9) and the rest use she/her pronouns (n=4). 
One participant was based in the UK while the rest were spread across eight US states.
Participants self-reported their ethnicities (Caucasian/white, Asian, and Black/African, Hispanic/Latinx), represented a diverse range of ages (20--50+) and had a variety of educational backgrounds (high school through to undergraduate, graduate, and trade school).
Nine participants self-reported as slightly or moderately familiar with statistical concepts and data visualization methods, two as expertly familiar, and one as not at all familiar. 
Five participants described data analysis and visualization tools as an important component in their professional workflows, and 8 interacted with data or visualizations more than 1--2 times/week.

\subsection{Quantitative Results}
\bgroup
\begin{table*}[]
\caption{
Rating scores for each prototype (Table, Multi-view, Targeted) on a five point Likert scale where {1} = Very Difficult (Very Unenjoyable) and {5} = Very Easy (Very Enjoyable).
Median scores are shown in boldface, averages in brackets, standard deviations in parentheses.
}
\vspace{-2mm}

\centering
\small
\def\arraystretch{1}

\begin{tabular}{m{7cm} m{1.5cm} c c c}
\hline
\textsc{prompt} : \emph{When using this prototype ...}
& \textsc{task}~\cite{brehmer_multi-level_2013}
& \textsc{table} 
& \textsc{multi-view}
& \textsc{targeted}
\\
\hline

\rowcolor{anti-flashwhite}
\emph{How enjoyable was it to interact with the data?}
& \texttt{enjoy}
& \textbf{3} $[3.31]$ $(0.95)$
& \textbf{4} $[3.77]$ $(1.01)$
& \textbf{4} $[3.54]$ $(0.97)$
\\

\emph{How easy was it to generate and answer questions?}
& \texttt{discover}
& \textbf{4} $[3.15]$ $(1.34)$
& \textbf{3} $[3.00]$ $(1.08)$
& \textbf{3} $[3.23]$ $(1.17)$

\\

\rowcolor{anti-flashwhite}
\emph{If you already knew what information you were trying to find, how easy would it be to look up or locate those data?}
& \texttt{lookup-} \texttt{locate}
& \textbf{3} $[3.31]$ $(1.32)$
& \textbf{4} $[3.77]$ $(1.17)$
& \textbf{4} $[3.38]$ $(1.19)$
\\

\emph{If you didn't already know which information you were trying to find, how easy would it be to browse or explore the data?}
& \texttt{browse-} \texttt{explore}
& \textbf{2} $[3.00]$ $(1.68)$
& \textbf{2} $[2.69]$ $(1.11)$
& \textbf{3} $[3.00]$ $(1.29)$
\\

\hline
\textsc{prompt} : \emph{When using this prototype ...}
& \textsc{use~\cite{karahanna_psychological_1999}}
& \textsc{table} 
& \textsc{multi-view}
& \textsc{targeted}
\\
\hline

\rowcolor{anti-flashwhite}
\emph{How easy was it to learn to use?}
& ease-of-use 
& \textbf{4} $[4.15]$ $(0.99)$
& \textbf{3} $[2.69]$ $(0.75)$
& \textbf{3} $[3.15]$ $(1.34)$
\\

\emph{How useful would it be to have access to this interaction style for engaging with data?}
& perceived usefulness 
& \textbf{4} $[4.15]$ $(0.80)$
& \textbf{4} $[4.00]$ $(0.82)$
& \textbf{4} $[4.15]$ $(1.07)$
\\

\hline
\end{tabular}
\label{table:likert-ratings}
\vspace{-5mm}
\end{table*}
\egroup


To supplement the cooperative interpretation sessions, participants rated each prototype using a series of Likert questions.
We designed a questionnaire with six prompts measuring a subset of Brehmer and Munzner's multi-level typology of abstract visualization tasks~\cite{brehmer_multi-level_2013}.
This framework, however, required some adaptation for non-visual modes of search.
In particular, searching with a screen reader requires a sequential approach to data that is at odds with the ``at-a-glance'' approach sighted readers take to browsing and exploring data.
As our prototypes focus on navigation through charts, we collapsed the \emph{location} dimension of Brehmer and Munzner's search decomposition resulting in two prompts that jointly measure \texttt{lookup-locate} and \texttt{browse-explore}.
We formulated additional questions to measure Brehmer and Munzner's \texttt{discover} and \texttt{enjoy} tasks as well as more traditional aspects of technology acceptance including \emph{ease-of-use} and \emph{perceived usefulness}~\cite{karahanna_psychological_1999}.
Participants responded on a five point scale where {1 = Very Difficult/Unenjoyable} and {5 = Very Easy/Enjoyable}.

Table~\ref{table:likert-ratings} displays the questionnaire prompts, their corresponding tasks, and statistics summarizing the participants' ratings.
A Friedman test
found a significant rating difference for the ease-of-use of the prototypes $\chi^2(2, N=13)=15.05, p<0.01$, with a large effect size (Kendall's $W=0.58$).
Follow-up Nemenyi tests
revealed that \Multiview{} was more difficult to use than \Table{} with statistical significance $(p<0.01)$, but \Target{} was not.
Additional tests for the other prompts found neither statistically significant differences, nor large effect sizes, between the prototypes.
However, median scores (which are more robust to outliers than means~\cite{morris_rich_2018}) suggest that participants generally \texttt{enjoy} interacting with \Multiview{} and \Target{} more, and found them easier to \texttt{lookup} or \texttt{locate} data with.
Moreover, \Target{} had the highest median score for affording \texttt{browse} or \texttt{explore} capabilities.
Conversely \Table{} was easiest to learn to use, and generally made it easy to \texttt{discover}, or ask and answer questions about the data.
Notably, in response to the question \say{\emph{How useful would it be to have access to this interaction style
for engaging with data?}} participants on average ranked all prototypes as more-than-useful $(med=4, \mu\geq 4)$.
These statistics provide only a partial picture of participants' experiences with the prototypes~\cite{bagozzi_legacy_2007}. 
Thus, we elucidate and contextualize reasons behind their scores through qualitative analysis.

\subsection{Qualitative Results}

After the interviews, we qualitatively coded the notes taken by the logger with a grounded theory approach \cite{charmaz_constructing_2006}. We performed open coding in parallel with the interviews (i.e., coding Monday's interviews after finishing Tuesday's interviews). We then synthesized the codes into memos, from which we derived these themes.

\textbf{Tables are familiar, tedious, but necessary.} Every participant noted that tables were their primary way of accessing data and visualizations. While tables are an important accessible option, participants overwhelmingly reported the same problems: they are ill-suited for processing large amounts of data and impose high cognitive load as users must remember previous lines of the table in order to contextualize subsequent values. 
As P2 reported, \hlquote{if I'm trying to get a general sense of the table, I'll just scroll through and see what values there are. But there's 393 rows, so I'll never scroll through all of it...I can't really get a snapshot.} 
P11 said that \hlquote{Finding relationships can be tricky if you're in a table, because you've got to either have a really good memory or just get really lucky. [...] If I didn't know what I was looking for, forget it.} 
At most, participants tabbed through 20--30 rows during our sessions, but did so only because of the questions we posed (e.g., ``is there a relationship between horsepower and mileage?'') and noted that if they encountered this table outside of the study, they would tab past a few rows to check for summary statistics and then move on. 

While it is not enjoyable to explore or build a mental model of data with static tables, participants still emphasized their necessity because of the format's familiarity: \hlquote{in terms of accessibility, tables are infinitely more useful because there is a standard way of navigating them in whatever your preferred screen reader is. With different representations, a blind person may not be trained to interpret it} (P2). 
This builds on prior literature \cite{sharif_understanding_2021} and echoes testimony from participants who had some difficulty with the new prototypes; they reported that they lacked expertise and therefore found it difficult to work with non-tabular data (P8, 10). In other words, to maximize accessibility, it is crucial to include a table view of the data \textit{in addition to} other forms of novel interaction. 


\textbf{Prior exposure to data analysis and representations increases the efficacy of spatial representations.} 
Participants who had experience conducting data analysis or reading tactile graphs/maps were able to easily develop a spatial understanding of how each prototype worked. 
Five participants (P2--4, 11, 13) made direct connections between the \Multiview{} and \Target{} prototypes, and the tactile graphs they encountered in school. 
Three participants (P2, 11, 12) found their software engineering experience made it easier to understand and navigate the prototypes' hierarchical structure. 
Previous literature on tactile mapping has also shown how developing tactile graphical literacy is crucial for building spatial knowledge, but they emphasize that it is not a sufficient for being able to conduct and understand data analysis. \cite{hasty_guidelines_2011,godfrey_advice_2015}
Since our participants already had an existing spatial framework, it became easier to explain how a prototype might work using their prior experience as a benchmark, which has been corroborated by similar studies in tactile cartography. \cite{wiedel_tactual_1969,aldrich_tactile_2001,sheppard_tactile_2001}
Importantly, our participants were able to find specific origin points that they could return to in order to navigate the different branches of the tree, which would be further aided with help menus and mini-tutorials to understand the keyboard shortcuts (P2). 
Being able to shift between origin points is especially important for switching between graphs or between variables.
By contrast, participants who had more difficulty with the prototypes (P8, 10) pointed to their lack of experience working with non-tabular data. P10 reported that being able to mentally visualize data points within a grid was a specific challenge.
\hlquote{I suspect that this might be understandable to someone who's done this before,} he said, \hlquote{I don't do well with these charts unless they're converted back into tables.}

\textbf{\textit{Structure}: Hierarchical representations make it possible to effectively convey insights with minimal cognitive load.} 
While static tables are the most common accessible option to interactive visualizations, eight of our participants (P2--5, 7, 10, 11, 13) expressed a desire to filter and sort the data so that they could begin to explore possible trends without wading line by line. 
Sorting and filtering a table is one way to look for trends but, to get a summary view of the data quickly, a system must provide snapshots in smaller intervals so that users  can easily construct a larger picture or choose specific slices of the data to explore further (i.e., ``details on demand''). \cite{shneiderman_eyes_2003,kim_accessible_2021} 
With \Multiview{} and \Target{} P4 said, \hlquote{I always want more layers and details, but some charts had too much...This was a happy medium between having the information I wanted and presenting it in a way that I can keep up with.} P5 also noted that he liked \hlquote{having the ability to scroll through at a higher level and then drill down deeper if that's of interest.} By giving users a way to quickly skip through the data across specific axes, they are able to rapidly generate a broader mental image of each graph and drill down further to collect more details. \hlquote{When I was working with the table, I [started building] a table in my head,} P2 shared. \hlquote{I had a rough representation of it as a scatter plot. But here, I know how to drill down and up between different layers of data grids, so that I can get the overall picture... 
[It gives me] different ways of thinking.}
Being able to control the parts of the data that were most important to them was also an issue of trust, as it also provided a way for users to reach conclusions for themselves rather than rely on the interpretation of others: \hlquote{It's hard to mix...doing your own analysis and be given a text description that you have to just trust} (P12). 
In their own workflows, these participants reported downloading static tables to further examine and manipulate with Excel, which they would use to create summary statistics or intervals to move more quickly through the data.


\textbf{\textit{Navigation}: Reading a visualization with a screen reader entails constant hypothesis testing and pattern-making.}
Since screen reader users parse data iteratively, nine of our participants (P1--5, 7, 8, 11, 13) described reading a visualization as a process of slowly building up a mental model and constantly testing it to see where the patterns may no longer hold. 
\hlquote{I'm going row by row, not memorizing exact numbers but building a pattern in my head, and looking at the other rows to test my theory,} reported P3. 
In other words, our participants 
engaged in a continuous state of updating and validating \cite{munzner_nested_2009} their mental images as new data challenged the existing patterns they have pieced together.  
\Multiview{} and \Target{} accelerated this process, as participants were able to more rapidly identify specific components that they wanted to test. 
For example, P2 intentionally moved quickly across each level of the structure hoping to find its ``edges,'' or the minimum and maximum limits of each axis and grid. 
\hlquote{Visually, it might look like I'm doing a lot of jumping around,} he said, \hlquote{[but it's] because I'm trying to build the picture in a way that makes sense for me.} 
Similarly, P5 started building his mental model of the visualization by drilling up and down the grid to create a spatial image of the data: 
\hlquote{I'm thinking more in spatial terms just because [this] is a new method of navigating to me. [...] I'm moving through the grid...I'm thinking of drilling down into that square to get more information.} 



\Target{} made it especially easy for participants to test their hypotheses by giving them direct access to components that might break their hypotheses. 
P5 reported that it allowed him to \hlquote{navigate to areas...that I'm interested in, skipping over stuff that's not of interest,} and P4 likened it to \hlquote{[being] able to go directly to what you want in a grocery inventory rather than going through each item one by one.} 
The ability to use structural, spatial, and target navigation in both \Multiview{} and \Target{} respectively facilitated the hypothesis-testing and pattern-making behaviors that our participants were accustomed to with static tables, and gave them an additional mental model for working with the data. 
As P1 noted, these prototypes gave her a richer understanding of the data by helping her piece together \hlquote{both the picture and the mathematical pattern,} whereas \Table{} afforded only the latter.



\textbf{\textit{Description}: Cursors and roadmaps are important for understanding where you are.} 
Being able to capture both a high-level overview of the information while preserving the ability to drill down into the data is a crucial component to accessing interactive visualizations~\cite{sharif_understanding_2021}. 
To navigate between these two levels, however, our participants emphasized the importance of markers to help them understand where they could move. 
\Target{} addressed this with dropdown menus that allowed participants to navigate to any part of the visualization, explore, and then return to where they had started. 
In the words of P4, \hlquote{[This] mode is freedom for the user. Being able to jump around and move in real time as you would with your hand gives you a new way of exploring the information.}
\Multiview{} approached this issue by allowing participants to move throughout the grid.
\hlquote{With the table, I was trying to hold the numbers in my head and I wasn't trying to visualize it or anything,} said P3. \hlquote{With [\textsc{Multi-View}], I can sort of think about it more like a visualization since I can move up and down, left and right. Even though I can use the arrows in the table, it just doesn't feel the same. I'm still feeling around and seeing what I can find.}
Without these navigation tools, P7 noted that \hlquote{It's too easy to get lost 
...I don't know how to backtrack.}
To orient herself, P13 would first test to see if she was at the corner cells in the visualizations (e.g., \hlquote{Am I in the upper left or the bottom right cell here?}) so that she could contextualize her position within the visualization and return to a point of origin.
\hlquote{I know that I must be at the bottom left cell here because I can't go to the left,} P13 said, \hlquote{but being able to know where that is beforehand would be very helpful.}

\section{Discussion and Future Work}
\label{section:Discussion}

In this paper, we explore how structure, navigation, and description compose together to yield richer screen reader experiences for data visualizations than are possible via alt text, data tables, or the current ARIA specification. 
Our results suggest promising next steps about accessible interaction and representation for visualizations.


\subsection{Enabling Richer Screen Reader Experiences Today}

Although our design dimensions highlight a diverse landscape of screen reader experiences for data visualizations, our study participants attested to the value of following existing best practices.
Namely, alt text and data tables provide a good baseline for making visualizations accessible.
Thus, visualization authors should consider adopting our design dimensions to enable more granular information access patterns only after these initial pieces are in place.

Existing visualization authoring methods, however, are likely insufficient for instantiating our design dimensions or producing usable experiences for screen reader users. 
In particular, it currently falls entirely on visualization authors to handcraft appropriate structures, navigational techniques, and description schemes on a per-visualization basis.
As a result, besides being a time-consuming endeavor, idiosyncratic implementations can introduce friction to the reading process.
For instance, per-visualization approaches might not account for an individual user's preferences in terms of verbosity, speed, or order of narrated output\,---\,three properties which varied widely among our study participant in ways that did not correlate with education level or experience with data. 
Thus, to scale and standardize this process, some responsibility for making visualizations screen reader accessible must be shared by \emph{toolkits} as well.
For example, our prototypes suggest a strategy for translating visualization specifications into hierarchical encoding structures (i.e., encoding channels as individual branches, and using visual affordances such as axis ticks and grid lines to populate the hierarchy levels).
If toolkits provide default experiences out-of-the-box, visualization authors can instead focus on customizing them to be more meaningful for their specific visualization, and screen reader users have a stronger guarantee that the resultant experiences will be more usable and respectful of their individual preferences. 

Current web accessibility standards also present limitations for realizing our design dimensions. 
For instance, there is no standard way to determine which element the screen reader cursor is selecting. 
Where ARIA has thus far focused on annotating documents with the semantics of a pre-defined palette of widgets, future web standards might instead express how elements respond to the \emph{interaction affordances} of screen readers.
For example, ARIA could offer explicit support for overview/detail hierarchies and different levels of description detail that can be progressively read according to user preferences.

\subsection{Studying and Refining the Design Dimensions} 

Our conversations with study participants also helped highlight that design considerations can differ substantially for users who are totally blind compared to those who have low-vision.
For example, partially-sighted participants used screen magnifiers alongside screen readers.
As a result, they preferred verbose written descriptions alongside more terse verbal narration.
Magnifier users also wished for in situ tooltips, which would eliminate the need to scroll back and forth between points and axes to understand data values.
However, promisingly, we found that using a screen reader and magnifier together affords unique benefits: \hlquote{I would have missed this point visually if I solely relied on the magnifier because the point is hidden behind another point} (P12).
Future work should more deeply explore how accommodations might complement and conflict when designing for different kinds of visual disability.

Similarly, in scoping our focus to screen readers and, thus, text-to-speech narration, we refrained from considering multi-sensory modalities in our design dimensions.
Yet, we found that most participants had previous experience with multi-sensory visualization, including sonification (P5, 7, 9, 13), tactile statistical charts (P2--4, 10, 11, 13), and haptic graphics (P3, 4, 11, 13). 
Some participants reported that a \textit{combination} of modalities would further enhance their experience\,---\,for example, getting a sonic overview of a line chart before reading more detailed text descriptions.
Other participants, however, cautioned that adding multiple modalities can create additional confusion.
For example, P7 noted that \hlquote{There's often a lack of explanation about how to map between sound and text.}
Based on this testimony, it is unlikely that ``sensory modalities'' are merely an additional, independent dimension within our framework.
Rather, future work must unpack the affordances of individual modalities, how they interact with one another, and how they impact the design of structure, navigation, and description. 


\vspace{-3mm}
\subsection{What are Accessible Interactions for Data Visualizations?} 

In visualization research, we typically distinguish between static and interactive visualizations, where the latter allows readers to actively manipulate visualized elements or the backing data pipeline~\cite{yi_toward_2007, heer_interactive_2012}.
Screen readers, however, complicate this model: reading is no longer a process that occurs purely ``in the head'' but rather becomes an embodied and interactive experience, as screen reader users must intentionally perform actions with their input devices in order to step through the visualization structure. 
While some aspects of this dichotomy may still hold, it is unclear how to cleanly separate \textit{static reading} from \textit{interactive manipulation} in the context of screen reader accessible visualizations, if these notions are conceptually separable at all.
For instance, \censor{Hajas} likened the navigation our prototypes afforded to \dhquote{shifting eye gaze, shifting focus of perceptual attention. When I navigate a visualization, naturally I would say `I'm looking at this figure' and not that `I'm interacting with this figure'.} Analogously, recent results in graphical perception find that sighted readers do not simply ``see'' visualizations in a single glance but rather perform active visual filtering operations~\cite{boger_jurassic_2021}.
However, when using the binary tree prototype (Fig.~\ref{fig:example-gallery}e), \censor{Hajas} noted a more distinct shift from reading to interacting. He said, \dhquote{it gave me the impression that I'm not just looking selectively, but I focus and zoom into the data,} analogous to zoom interactions that change the viewport for sighted readers.
Better characterizing the shift that occurs with this prototype, and exploring accessible manipulations of visualizations that allow screen reader users to meaningfully conduct data analysis, are compelling opportunities for future work.

\section*{Acknowledgements}
We thank Matthew Blanco, Evan Peck, the NYU Digital Theory Lab, and the  MIT Accessibility Office. This work was supported by NSF awards \#1942659, \#1122374, and \#1941577.
\printbibliography 

@techreport{amick_guidelines_1997,
	title = {Guidelines for {Design} of {Tactile} {Graphics}},
	url = {https://www.aph.org/research/guides/},
	urldate = {2019-06-13},
	institution = {American Printing House for the Blind},
	author = {Amick, Nancy and Corcoran, Jane},
	year = {1997},
}

@article{aldrich_tactile_2001,
	title = {Tactile {Graphics} {In} {School} {Education}: {Perspectives} {From} {Pupils}},
	volume = {19},
	issn = {0264-6196},
	shorttitle = {Tactile graphics in school education},
	url = {https://doi.org/10.1177/026461960101900204},
	doi = {10.1177/026461960101900204},
	number = {2},
	urldate = {2019-06-13},
	journal = {British Journal of Visual Impairment},
	author = {Aldrich, Frances K. and Sheppard, Linda},
	year = {2001},
	pages = {69--73},
}

@article{sheppard_tactile_2001,
	title = {Tactile {Graphics} {In} {School} {Education}: {Perspectives} {From} {Pupils}},
	volume = {19},
	issn = {0264-6196},
	shorttitle = {Tactile graphics in school education},
	url = {https://doi.org/10.1177/026461960101900303},
	doi = {10.1177/026461960101900303},
	language = {en},
	number = {3},
	journal = {British Journal of Visual Impairment},
	author = {Sheppard, Linda and Aldrich, Frances K.},
	year = {2001},
	pages = {93--97},
}

@inproceedings{li_editing_2019,
	address = {Glasgow, Scotland Uk},
	title = {Editing {Spatial} {Layouts} through {Tactile} {Templates} for {People} with {Visual} {Impairments}},
	isbn = {978-1-4503-5970-2},
	url = {http://dl.acm.org/citation.cfm?doid=3290605.3300436},
	doi = {10.1145/3290605.3300436},
	language = {en},
	urldate = {2019-05-31},
	booktitle = {{ACM} {Conference} on {Human} {Factors} in {Computing} {Systems} ({CHI})},
	author = {Li, Jingyi and Kim, Son and Miele, Joshua A. and Agrawala, Maneesh and Follmer, Sean},
	year = {2019},
	pages = {1--11},
}

@article{goos_haptic_2001,
	title = {Haptic {Graphs} {For} {Blind} {Computer} {Users}},
	volume = {2058},
	url = {http://link.springer.com/10.1007/3-540-44589-7_5},
	doi = {10.1007/3-540-44589-7_5},
	language = {en},
	urldate = {2019-06-11},
	journal = {Lecture Notes in Computer Science},
	author = {Yu, Wai and Ramloll, Ramesh and Brewster, Stephen},
	editor = {Brewster, Stephen and Murray-Smith, Roderick},
	collaborator = {Goos, Gerhard and Hartmanis, Juris and van Leeuwen, Jan},
	year = {2001},
	pages = {41--51},
}

@article{kim_accessible_2021,
	title = {Accessible {Visualization}: {Design} {Space}, {Opportunities}, and {Challenges}},
	volume = {40},
	issn = {0167-7055, 1467-8659},
	shorttitle = {Accessible {Visualization}},
	url = {https://onlinelibrary.wiley.com/doi/10.1111/cgf.14298},
	doi = {10.1111/cgf.14298},
	language = {en},
	number = {3},
	urldate = {2021-07-12},
	journal = {Computer Graphics Forum (EuroVis)},
	author = {Kim, N. W. and Joyner, S. C. and Riegelhuth, A. and Kim, Y.},
	year = {2021},
	pages = {173--188},
}

@incollection{butler_technology_2021,
	address = {New York, NY, USA},
	title = {Technology {Developments} in {Touch}-{Based} {Accessible} {Graphics}: {A} {Systematic} {Review} of {Research} 2010-2020},
	isbn = {978-1-4503-8096-6},
	shorttitle = {Technology {Developments} in {Touch}-{Based} {Accessible} {Graphics}},
	url = {https://doi.org/10.1145/3411764.3445207},
	number = {278},
	urldate = {2021-07-07},
	booktitle = {{ACM} {Conference} on {Human} {Factors} in {Computing} {Systems} ({CHI})},
	author = {Butler, Matthew and Holloway, Leona M and Reinders, Samuel and Goncu, Cagatay and Marriott, Kim},
	year = {2021},
	pages = {1--15},
}

@inproceedings{de_greef_interdependent_2021,
	address = {New York, NY, USA},
	series = {{ASSETS} '21},
	title = {Interdependent {Variables}: {Remotely} {Designing} {Tactile} {Graphics} for an {Accessible} {Workflow}},
	isbn = {978-1-4503-8306-6},
	shorttitle = {Interdependent {Variables}},
	url = {https://doi.org/10.1145/3441852.3476468},
	doi = {10.1145/3441852.3476468},
	urldate = {2021-11-17},
	booktitle = {{ACM} {Conference} on {Computers} and {Accessibility} ({SIGACCESS})},
	publisher = {Association for Computing Machinery},
	author = {de Greef, Lilian and Moritz, Dominik and Bennett, Cynthia},
	year = {2021},
	pages = {1--6},
}

@inproceedings{satyanarayan_vega-lite_2017,
	title = {Vega-{Lite}: {A} {Grammar} of {Interactive} {Graphics}},
	shorttitle = {Vega-{Lite}},
	url = {https://ieeexplore.ieee.org/abstract/document/7539624},
	doi = {10.1109/TVCG.2016.2599030},
	booktitle = {{IEEE} {Transactions} on {Visualization} \& {Computer} {Graphics} ({Proc}. {IEEE} {VIS})},
	author = {Satyanarayan, Arvind and Moritz, Dominik and Wongsuphasawat, Kanit and Heer, Jeffrey},
	year = {2017},
}

@book{bertin_semiology_1983,
	title = {Semiology of {Graphics}},
	isbn = {978-0-299-09060-9},
	url = {https://dl.acm.org/doi/10.5555/1095597},
	publisher = {University of Wisconsin Press},
	author = {Bertin, Jacques},
	year = {1983},
}

@article{munzner_nested_2009,
	title = {A {Nested} {Model} for {Visualization} {Design} and {Validation}},
	volume = {15},
	issn = {1941-0506},
	url = {https://ieeexplore.ieee.org/abstract/document/5290695},
	doi = {10.1109/TVCG.2009.111},
	number = {6},
	journal = {IEEE Transactions on Visualization and Computer Graphics (Proc. IEEE VIS)},
	author = {Munzner, Tamara},
	year = {2009},
	note = {Conference Name: IEEE Transactions on Visualization and Computer Graphics},
	pages = {921--928},
}

@inproceedings{lundgard_accessible_2021,
	title = {Accessible {Visualization} via {Natural} {Language} {Descriptions}: {A} {Four}-{Level} {Model} of {Semantic} {Content}},
	url = {http://vis.csail.mit.edu/pubs/vis-text-model},
	doi = {10.1109/TVCG.2021.3114770},
	booktitle = {{IEEE} {Transactions} on {Visualization} \& {Computer} {Graphics} ({Proc}. {IEEE} {VIS})},
	author = {Lundgard, Alan and Satyanarayan, Arvind},
	year = {2021},
	pages = {11},
}

@techreport{gould_effective_2008,
	title = {Effective {Practices} for {Description} of {Science} {Content} within {Digital} {Talking} {Books}},
	url = {https://www.wgbh.org/foundation/ncam/guidelines/effective-practices-for-description-of-science-content-within-digital-talking-books},
	language = {en},
	urldate = {2020-11-03},
	institution = {The WGBH National Center for Accessible Media},
	author = {Gould, Bryan and O’Connell, Trisha and Freed, Geoffrey},
	year = {2008},
}

@inproceedings{potluri_examining_2021,
	title = {Examining {Visual} {Semantic} {Understanding} in {Blind} and {Low}-{Vision} {Technology} {Users}},
	url = {https://dl.acm.org/doi/abs/10.1145/3411764.3445040},
	doi = {10.1145/3411764.3445040},
	booktitle = {{ACM} {Conference} on {Human} {Factors} in {Computing} {Systems} ({CHI})},
	author = {Potluri, Venkatesh and Grindeland, Tadashi E and Froehlich, Jon E. and Mankoff, Jennifer},
	year = {2021},
}

@techreport{hasty_guidelines_2011,
	title = {Guidelines and {Standards} for {Tactile} {Graphics}},
	url = {http://www.brailleauthority.org/tg/},
	institution = {Braille Authority of North America},
	author = {Hasty, Lucia and Milbury, Janet and Miller, Irene and O'Day, Allison and Acquinas, Pather and Spence, Diance},
	year = {2011},
}

@inproceedings{wu_understanding_2021,
	title = {Understanding {Data} {Accessibility} for {People} with {Intellectual} and {Developmental} {Disabilities}},
	url = {https://dl.acm.org/doi/abs/10.1145/3411764.3445743},
	doi = {10.1145/3411764.3445743},
	language = {en},
	booktitle = {{ACM} {Conference} on {Human} {Factors} in {Computing} {Systems} ({CHI})},
	author = {Wu, Keke and Petersen, Emma and Ahmad, Tahmina and Burlinson, David and Tanis, Shea and Szafir, Danielle Albers},
	year = {2021},
}

@inproceedings{lundgard_sociotechnical_2019,
	title = {Sociotechnical {Considerations} for {Accessible} {Visualization} {Design}},
	url = {https://ieeexplore.ieee.org/abstract/document/8933762},
	doi = {10.1109/VISUAL.2019.8933762},
	booktitle = {{IEEE} {Transactions} on {Visualization} \& {Computer} {Graphics} ({Proc}. {IEEE} {VIS})},
	author = {Lundgard, Alan and Lee, Crystal and Satyanarayan, Arvind},
	year = {2019},
}

@inproceedings{morris_rich_2018,
	address = {Montreal QC Canada},
	title = {Rich {Representations} of {Visual} {Content} for {Screen} {Reader} {Users}},
	isbn = {978-1-4503-5620-6},
	url = {https://dl.acm.org/doi/10.1145/3173574.3173633},
	doi = {10.1145/3173574.3173633},
	language = {en},
	urldate = {2021-08-08},
	booktitle = {{ACM} {Conference} on {Human} {Factors} in {Computing} {Systems} ({CHI})},
	author = {Morris, Meredith Ringel and Johnson, Jazette and Bennett, Cynthia L. and Cutrell, Edward},
	year = {2018},
	pages = {1--11},
}

@article{choi_visualizing_2019,
	title = {Visualizing for the {Non}‐{Visual}: {Enabling} the {Visually} {Impaired} to {Use} {Visualization}},
	volume = {38},
	issn = {0167-7055, 1467-8659},
	shorttitle = {Visualizing for the {Non}‐{Visual}},
	url = {https://onlinelibrary.wiley.com/doi/10.1111/cgf.13686},
	doi = {10.1111/cgf.13686},
	language = {en},
	number = {3},
	urldate = {2021-07-24},
	journal = {Computer Graphics Forum (EuroVis)},
	author = {Choi, Jinho and Jung, Sanghun and Park, Deok Gun and Choo, Jaegul and Elmqvist, Niklas},
	month = jun,
	year = {2019},
	pages = {249--260},
}

@article{godfrey_advice_2015,
	title = {Advice {From} {Blind} {Teachers} on {How} to {Teach} {Statistics} to {Blind} {Students}},
	volume = {23},
	issn = {null},
	url = {https://doi.org/10.1080/10691898.2015.11889746},
	doi = {10.1080/10691898.2015.11889746},
	number = {3},
	urldate = {2021-07-20},
	journal = {Journal of Statistics Education},
	author = {Godfrey, A. Jonathan R. and Loots, M. Theodor},
	year = {2015},
	pages = {null},
}

@incollection{miesenberger_universal_2018,
	address = {Cham},
	title = {Universal {Design} {Tactile} {Graphics} {Production} {System} {BPLOT4} for {Blind} {Teachers} and {Blind} {Staffs} to {Produce} {Tactile} {Graphics} and {Ink} {Print} {Graphics} of {High} {Quality}},
	volume = {10897},
	isbn = {978-3-319-94273-5 978-3-319-94274-2},
	url = {http://link.springer.com/10.1007/978-3-319-94274-2_23},
	language = {en},
	urldate = {2021-07-23},
	booktitle = {Computers {Helping} {People} with {Special} {Needs}},
	publisher = {Springer International Publishing},
	author = {Fujiyoshi, Mamoru and Fujiyoshi, Akio and Tanaka, Hiroshi and Ishida, Toru},
	editor = {Miesenberger, Klaus and Kouroupetroglou, Georgios},
	year = {2018},
	pages = {167--176},
}

@article{baker_tactile_2016,
	title = {Tactile {Graphics} with a {Voice}},
	volume = {8},
	issn = {1936-7228},
	url = {http://doi.org/10.1145/2854005},
	doi = {10.1145/2854005},
	number = {1},
	urldate = {2021-07-23},
	journal = {ACM Transactions on Accessible Computing (TACCESS)},
	author = {Baker, Catherine M. and Milne, Lauren R. and Drapeau, Ryan and Scofield, Jeffrey and Bennett, Cynthia L. and Ladner, Richard E.},
	year = {2016},
	pages = {3:1--3:22},
}

@article{weninger_asvg_2015,
	title = {{ASVG} {Accessible} {Scalable} {Vector} {Graphics}: {Intention} {Trees} {To} {Make} {Charts} {More} {Accessible} {And} {Usable}},
	volume = {9},
	shorttitle = {{ASVG} − {Accessible} {Scalable} {Vector} {Graphics}},
	doi = {10.1108/JAT-10-2015-0124},
	journal = {Journal of Assistive Technologies},
	author = {Weninger, Markus and Ortner, Gerald and Hahn, Tobias and Druemmer, Olaf and Miesenberger, Klaus},
	year = {2015},
	pages = {239--246},
}

@inproceedings{brock_usage_2010,
	address = {New York, NY, USA},
	series = {{ITS} '10},
	title = {Usage {Of} {Multimodal} {Maps} {For} {Blind} {People}: {Why} {And} {How}},
	isbn = {978-1-4503-0399-6},
	shorttitle = {Usage of multimodal maps for blind people},
	url = {https://doi.org/10.1145/1936652.1936699},
	doi = {10.1145/1936652.1936699},
	urldate = {2021-07-23},
	booktitle = {{ACM} {International} {Conference} on {Interactive} {Tabletops} and {Surfaces} ({ISS})},
	author = {Brock, Anke and Truillet, Philippe and Oriola, Bernard and Jouffrais, Christophe},
	year = {2010},
	pages = {247--248},
}

@article{cornwall_what_1995,
	title = {What {Is} {Participatory} {Research}?},
	volume = {41},
	issn = {0277-9536},
	url = {https://www.sciencedirect.com/science/article/pii/027795369500127S},
	doi = {10.1016/0277-9536(95)00127-S},
	language = {en},
	number = {12},
	urldate = {2022-03-09},
	journal = {Social Science \& Medicine},
	author = {Cornwall, Andrea and Jewkes, Rachel},
	year = {1995},
	pages = {1667--1676},
}

@article{saunders_anonymising_2015,
	title = {Anonymising {Interview} {Data}: {Challenges} {And} {Compromise} {In} {Practice}},
	volume = {15},
	issn = {1468-7941},
	shorttitle = {Anonymising interview data},
	url = {https://doi.org/10.1177/1468794114550439},
	doi = {10.1177/1468794114550439},
	language = {en},
	number = {5},
	urldate = {2022-03-06},
	journal = {Qualitative Research},
	author = {Saunders, Benjamin and Kitzinger, Jenny and Kitzinger, Celia},
	year = {2015},
	pages = {616--632},
}

@article{shew_ableism_2020,
	title = {Ableism, {Technoableism}, and {Future} {AI}},
	volume = {39},
	issn = {1937-416X},
	doi = {10.1109/MTS.2020.2967492},
	number = {1},
	journal = {IEEE Technology and Society Magazine},
	author = {Shew, Ashley},
	month = mar,
	year = {2020},
	pages = {40--85},
}

@misc{jackson_disability_2019,
	type = {Tweet},
	title = {Disability dongle: {A} well-intended, elegant, yet useless solution to a problem we never knew we had. {Disability} dongles are most often conceived of and created in design schools and at {IDEO}.},
	url = {https://twitter.com/elizejackson/status/1110629818234818570},
	journal = {Twitter},
	author = {Jackson, Liz},
	year = {2019},
}

@article{hamraie_designing_2013,
	title = {Designing {Collective} {Access}: {A} {Feminist} {Disability} {Theory} of {Universal} {Design}},
	volume = {33},
	issn = {2159-8371, 1041-5718},
	shorttitle = {Designing {Collective} {Access}},
	url = {http://dsq-sds.org/article/view/3871},
	doi = {10.18061/dsq.v33i4.3871},
	language = {en},
	number = {4},
	urldate = {2019-05-22},
	journal = {Disability Studies Quarterly},
	author = {Hamraie, Aimi},
	year = {2013},
}

@inproceedings{sengers_reflective_2005,
	address = {New York, NY, USA},
	series = {{CC} '05},
	title = {Reflective {Design}},
	isbn = {978-1-59593-203-7},
	url = {http://doi.acm.org/10.1145/1094562.1094569},
	doi = {10.1145/1094562.1094569},
	urldate = {2019-05-30},
	booktitle = {{ACM} {Conference} on {Critical} {Computing}: {Between} {Sense} and {Sensibility}},
	author = {Sengers, Phoebe and Boehner, Kirsten and David, Shay and Kaye, Joseph 'Jofish'},
	year = {2005},
	pages = {49--58},
}

@book{costanza-chock_design_2020,
	address = {Cambridge, MA},
	title = {Design {Justice}: {Towards} an {Intersectional} {Feminist} {Framework} for {Design} {Theory} and {Practice}},
	shorttitle = {Design {Justice}},
	url = {https://papers.ssrn.com/abstract=3189696},
	language = {en},
	urldate = {2019-07-25},
	publisher = {MIT Press},
	author = {Costanza-Chock, Sasha},
	year = {2020},
}

@misc{horst_palmerpenguins_2020,
	title = {Palmerpenguins: {Palmer} {Archipelago} ({Antarctica}) {Penguin} {Data}},
	copyright = {CC0-1.0},
	url = {https://github.com/allisonhorst/palmerpenguins},
	urldate = {2021-12-02},
	author = {Horst, Allison M. and Hill, Alison Presmanes and Gorman, Kristen B.},
	year = {2020},
}

@article{becker_visual_1996,
	title = {The {Visual} {Design} and {Control} of {Trellis} {Display}},
	volume = {5},
	issn = {1061-8600, 1537-2715},
	url = {http://www.tandfonline.com/doi/abs/10.1080/10618600.1996.10474701},
	doi = {10.1080/10618600.1996.10474701},
	language = {en},
	number = {2},
	urldate = {2021-12-02},
	journal = {Journal of Computational and Graphical Statistics},
	author = {Becker, Richard A. and Cleveland, William S. and Shyu, Ming-Jen},
	year = {1996},
	pages = {123--155},
}

@incollection{koch_state_2012,
	address = {Berlin, Heidelberg},
	series = {Lecture {Notes} in {Geoinformation} and {Cartography}},
	title = {State of the {Art} of {Tactile} {Maps} for {Visually} {Impaired} {People}},
	isbn = {978-3-642-12272-9},
	url = {https://doi.org/10.1007/978-3-642-12272-9_9},
	language = {en},
	urldate = {2021-07-20},
	booktitle = {True-{3D} in {Cartography}: {Autostereoscopic} and {Solid} {Visualisation} of {Geodata}},
	publisher = {Springer},
	author = {Koch, Wolf Günther},
	editor = {Buchroithner, Manfred},
	year = {2012},
	doi = {10.1007/978-3-642-12272-9_9},
	pages = {137--151},
}

@book{charmaz_constructing_2006,
	address = {London ; Thousand Oaks, Calif},
	title = {Constructing {Grounded} {Theory}},
	isbn = {978-0-7619-7352-2 978-0-7619-7353-9},
	url = {https://uk.sagepub.com/en-gb/eur/constructing-grounded-theory/book235960},
	language = {en},
	publisher = {Sage Publications},
	author = {Charmaz, Kathy},
	year = {2006},
}

@misc{sas_graphics_accelerator_sas_2018,
	title = {{SAS} {Graphics} {Accelerator} {Customer} {Product} {Page}},
	url = {https://support.sas.com/software/products/graphics-accelerator/index.html#s1=1},
	urldate = {2021-12-01},
	author = {SAS Graphics Accelerator},
	year = {2018},
}

@article{kaper_data_1999,
	title = {Data {Sonification} {And} {Sound} {Visualization}},
	volume = {1},
	issn = {1558-366X},
	doi = {10.1109/5992.774840},
	number = {4},
	journal = {Computing in Science Engineering},
	author = {Kaper, H.G. and Wiebel, E. and Tipei, S.},
	year = {1999},
	pages = {48--58},
}

@article{barrass_using_1999,
	title = {Using {Sonification}},
	volume = {7},
	issn = {1432-1882},
	url = {https://doi.org/10.1007/s005300050108},
	doi = {10.1007/s005300050108},
	language = {en},
	number = {1},
	urldate = {2021-12-01},
	journal = {Multimedia Systems},
	author = {Barrass, Stephen and Kramer, Gregory},
	year = {1999},
	pages = {23--31},
}

@inproceedings{boger_jurassic_2021,
	title = {Jurassic {Mark}: {Inattentional} {Blindness} for a {Datasaurus} {Reveals} that {Visualizations} are {Explored}, not {Seen}},
	shorttitle = {Jurassic {Mark}},
	url = {https://ieeexplore.ieee.org/abstract/document/9623273},
	doi = {10.1109/VIS49827.2021.9623273},
	language = {en},
	urldate = {2021-12-01},
	booktitle = {{IEEE} {Transactions} on {Visualization} \& {Computer} {Graphics} ({Proc}. {IEEE} {VIS})},
	author = {Boger, Tal and Most, Steven B. and Franconeri, Steven L.},
	year = {2021},
}

@misc{mdn_contributors_aria_2021,
	title = {{ARIA} - {Accessibility}},
	url = {https://developer.mozilla.org/en-US/docs/Web/Accessibility/ARIA},
	language = {en-US},
	urldate = {2021-12-01},
	journal = {MDN Web Docs},
	author = {{MDN Contributors}},
	month = nov,
	year = {2021},
}

@misc{w3c_wai-aria_2018,
	title = {{WAI}-{ARIA} {Graphics} {Module}},
	url = {https://www.w3.org/TR/graphics-aria-1.0/},
	urldate = {2021-12-01},
	author = {W3C},
	year = {2018},
}

@article{bagozzi_legacy_2007,
	title = {The {Legacy} of the {Technology} {Acceptance} {Model} and a {Proposal} for a {Paradigm} {Shift}.},
	volume = {8},
	issn = {1536-9323},
	url = {https://aisel.aisnet.org/jais/vol8/iss4/12},
	doi = {10.17705/1jais.00122},
	number = {4},
	journal = {Journal of the Association for Information Systems},
	author = {Bagozzi, Richard P.},
	year = {2007},
}

@misc{w3c_web_2018,
	title = {Web {Accessibility} {Laws} \& {Policies}},
	url = {https://www.w3.org/WAI/policies/},
	language = {en},
	urldate = {2021-11-30},
	journal = {Web Accessibility Initiative (WAI)},
	author = {{W3C}},
	year = {2018},
}

@inproceedings{martinez_accessible_2019,
	address = {New York, NY, USA},
	series = {Interacci\&\#xf3;n '19},
	title = {Accessible {Statistical} {Charts} {For} {People} {With} {Low} {Vision} {And} {Colour} {Vision} {Deficiency}},
	isbn = {978-1-4503-7176-6},
	url = {https://doi.org/10.1145/3335595.3335618},
	doi = {10.1145/3335595.3335618},
	booktitle = {{ACM} {International} {Conference} on {Human} {Computer} {Interaction}},
	author = {Martínez, Rubén Alcaraz and Turró, Mireia Ribera and Saltiveri, Toni Granollers},
	month = jun,
	year = {2019},
	pages = {1--2},
}

@misc{highcharts_accessibility_2021,
	title = {Accessibility {Module}},
	url = {https://highcharts.com/docs/accessibility/accessibility-module},
	language = {en},
	author = {Highcharts},
	year = {2021},
}

@inproceedings{brehmer_multi-level_2013,
	title = {A {Multi}-{Level} {Typology} of {Abstract} {Visualization} {Tasks}},
	url = {http://ieeexplore.ieee.org/document/6634168/},
	doi = {10.1109/TVCG.2013.124},
	language = {en},
	booktitle = {{IEEE} {Transactions} on {Visualization} \& {Computer} {Graphics} ({Proc}. {IEEE} {VIS})},
	author = {Brehmer, Matthew and Munzner, Tamara},
	year = {2013},
}

@article{karahanna_psychological_1999,
	title = {The {Psychological} {Origins} of {Perceived} {Usefulness} and {Ease}-of-use},
	volume = {35},
	issn = {0378-7206},
	url = {https://doi.org/10.1016/S0378-7206(98)00096-2},
	doi = {10.1016/S0378-7206(98)00096-2},
	number = {4},
	journal = {Information and Management},
	author = {Karahanna, Elena and Straub, Detmar W.},
	year = {1999},
	pages = {237--250},
}

@article{chundury_towards_2021,
	title = {Towards {Understanding} {Sensory} {Substitution} for {Accessible} {Visualization}: {An} {Interview} {Study}},
	issn = {1077-2626, 1941-0506, 2160-9306},
	shorttitle = {Towards {Understanding} {Sensory} {Substitution} for {Accessible} {Visualization}},
	url = {https://ieeexplore.ieee.org/document/9552177/},
	doi = {10.1109/TVCG.2021.3114829},
	language = {en},
	urldate = {2021-11-29},
	journal = {IEEE Transactions on Visualization and Computer Graphics (Proc. IEEE VIS)},
	author = {Chundury, Pramod and Patnaik, Biswaksen and Reyazuddin, Yasmin and Tang, Christine and Lazar, Jonathan and Elmqvist, Niklas},
	year = {2021},
	pages = {1--1},
}

@inproceedings{jung_communicating_2021,
	title = {Communicating {Visualizations} without {Visuals}: {Investigation} of {Visualization} {Alternative} {Text} for {People} with {Visual} {Impairments}},
	volume = {PP},
	shorttitle = {Communicating {Visualizations} without {Visuals}},
	url = {https://ieeexplore.ieee.org/abstract/document/9552938},
	doi = {10.1109/TVCG.2021.3114846},
	language = {eng},
	booktitle = {{IEEE} {Transactions} on {Visualization} and {Computer} {Graphics} ({Proc}. {IEEE} {VIS})},
	author = {Jung, Crescentia and Mehta, Shubham and Kulkarni, Atharva and Zhao, Yuhang and Kim, Yea-Seul},
	year = {2021},
	pmid = {34591768},
}

@misc{sarah_l_fossheim_how_2020,
	title = {How (not) to make accessible data visualizations, illustrated by the {US} presidential election.},
	url = {https://fossheim.io/writing/posts/accessible-dataviz-us-elections/},
	language = {en},
	urldate = {2021-05-20},
	author = {{Sarah L. Fossheim}},
	year = {2020},
}

@techreport{wiedel_tactual_1969,
	address = {Washington, D.C.},
	title = {Tactual {Mapping}: {Design}, {Reproduction}, {Reading} and {Interpretation}},
	url = {https://archive.org/details/tactualmappingde00jose_0},
	language = {en},
	number = {D-2557-S 1969},
	institution = {Department of Health, Education, and Welfare},
	author = {Wiedel, Joseph W. and Groves, Paul A.},
	year = {1969},
	pages = {140},
}

@incollection{shneiderman_eyes_2003,
	address = {San Francisco},
	series = {Interactive {Technologies}},
	title = {The {Eyes} {Have} {It}: {A} {Task} by {Data} {Type} {Taxonomy} for {Information} {Visualizations}},
	isbn = {978-1-55860-915-0},
	shorttitle = {The {Eyes} {Have} {It}},
	url = {https://www.sciencedirect.com/science/article/pii/B9781558609150500469},
	language = {en},
	urldate = {2021-11-28},
	booktitle = {The {Craft} of {Information} {Visualization}},
	publisher = {Morgan Kaufmann},
	author = {Shneiderman, Ben},
	editor = {Bederson, BENJAMIN B. and Shneiderman, BEN},
	year = {2003},
	doi = {10.1016/B978-155860915-0/50046-9},
	pages = {364--371},
}

@article{dimara_what_2020,
	title = {What is {Interaction} for {Data} {Visualization}?},
	volume = {26},
	url = {https://hal.archives-ouvertes.fr/hal-02197062},
	doi = {10.1109/TVCG.2019.2934283},
	number = {1},
	urldate = {2021-11-28},
	journal = {IEEE Transactions on Visualization and Computer Graphics (Proc. IEEE VIS)},
	author = {Dimara, Evanthia and Perin, Charles},
	year = {2020},
	pages = {119 -- 129},
}

@article{heer_interactive_2012,
	title = {Interactive {Dynamics} for {Visual} {Analysis}},
	volume = {55},
	issn = {0001-0782, 1557-7317},
	url = {https://dl.acm.org/doi/10.1145/2133806.2133821},
	doi = {10.1145/2133806.2133821},
	language = {en},
	number = {4},
	urldate = {2021-11-28},
	journal = {Communications of the ACM},
	author = {Heer, Jeffrey and Shneiderman, Ben},
	year = {2012},
	pages = {45--54},
}

@misc{freedom_scientific_jaws_2021,
	title = {{JAWS} {Web} {Verbosity}},
	url = {https://www.freedomscientific.com/SurfsUp/Web_Verbosity.htm},
	urldate = {2021-11-28},
	journal = {Freedom Scientific},
	author = {Freedom Scientific},
	year = {2021},
}

@misc{webaim_screen_2021,
	title = {Screen {Reader} {User} {Survey} \#9 {Results}},
	url = {https://webaim.org/projects/screenreadersurvey9/},
	urldate = {2021-11-27},
	journal = {WebAIM},
	author = {WebAIM},
	year = {2021},
}

@inproceedings{frokjaer_cooperative_2005,
	address = {New York, NY, USA},
	series = {{CHI} {EA} '05},
	title = {Cooperative {Usability} {Testing}: {Complementing} {Usability} {Tests} {With} {User}-{Supported} {Interpretation} {Sessions}},
	isbn = {978-1-59593-002-6},
	shorttitle = {Cooperative usability testing},
	url = {https://doi.org/10.1145/1056808.1056922},
	doi = {10.1145/1056808.1056922},
	urldate = {2021-11-27},
	booktitle = {{ACM} {Extended} {Abstracts} on {Human} {Factors} in {Computing} {Systems} ({CHI})},
	author = {Frøkjær, Erik and Hornbæk, Kasper},
	year = {2005},
	pages = {1383--1386},
}

@inproceedings{hutchinson_technology_2003,
	address = {New York, NY, USA},
	title = {Technology {Probes}: {Inspiring} {Design} for and with {Families}},
	isbn = {978-1-58113-630-2},
	shorttitle = {Technology {Probes}},
	url = {http://doi.acm.org/10.1145/642611.642616},
	doi = {10.1145/642611.642616},
	urldate = {2019-07-23},
	booktitle = {{ACM} {Conference} on {Human} {Factors} in {Computing} {Systems} ({CHI})},
	author = {Hutchinson, Hilary and Mackay, Wendy and Westerlund, Bo and Bederson, Benjamin B. and Druin, Allison and Plaisant, Catherine and Beaudouin-Lafon, Michel and Conversy, Stéphane and Evans, Helen and Hansen, Heiko and Roussel, Nicolas and Eiderbäck, Björn},
	year = {2003},
	pages = {17--24},
}

@inproceedings{sharif_understanding_2021,
	address = {New York, NY, USA},
	series = {{ASSETS} '21},
	title = {Understanding {Screen}-{Reader} {Users}’ {Experiences} with {Online} {Data} {Visualizations}},
	isbn = {978-1-4503-8306-6},
	url = {http://doi.org/10.1145/3441852.3471202},
	doi = {10.1145/3441852.3471202},
	urldate = {2021-11-27},
	booktitle = {{ACM} {Conference} on {Computers} and {Accessibility} ({SIGACCESS})},
	author = {Sharif, Ather and Chintalapati, Sanjana Shivani and Wobbrock, Jacob O. and Reinecke, Katharina},
	year = {2021},
	pages = {1--16},
}

@article{pirolli_information_1999,
	title = {Information {Foraging}},
	volume = {106},
	issn = {1939-1471(Electronic),0033-295X(Print)},
	doi = {10.1037/0033-295X.106.4.643},
	number = {4},
	journal = {Psychological Review},
	author = {Pirolli, Peter and Card, Stuart},
	year = {1999},
	pages = {643--675},
}

@article{higgins_supreme_2019,
	title = {Supreme {Court} {Hands} {Victory} {To} {Blind} {Man} {Who} {Sued} {Domino}'s {Over} {Site} {Accessibility}},
	url = {https://www.cnbc.com/2019/10/07/dominos-supreme-court.html},
	language = {en},
	urldate = {2021-11-27},
	journal = {CNBC},
	author = {Higgins, Tucker},
	year = {2019},
}

@misc{boxhall_accessibility_2022,
	title = {The {Accessibility} {Object} {Model} ({AOM})},
	url = {https://wicg.github.io/aom/},
	language = {en-US},
	urldate = {2021-11-27},
	journal = {AOM},
	author = {Boxhall, Alice and Craig, James and Mazzoni, Dominic and Surkov, Alexander},
	year = {2022},
}

@misc{davert_whats_2019,
	title = {What's {New} {In} {iOS} 13 {Accessibility} {For} {Individuals} {Who} {Are} {Blind} or {Deaf}-{Blind}},
	url = {https://www.applevis.com/blog/whats-new-ios-13-accessibility-individuals-who-are-blind-or-deaf-blind},
	urldate = {2021-11-18},
	journal = {AppleVis},
	author = {Davert, Scott and Editorial Team, AppleVis},
	year = {2019},
}

@article{yi_toward_2007,
	title = {Toward a {Deeper} {Understanding} of the {Role} of {Interaction} in {Information} {Visualization}},
	volume = {13},
	issn = {1941-0506},
	url = {https://ieeexplore.ieee.org/abstract/document/4376144},
	doi = {10.1109/TVCG.2007.70515},
	number = {6},
	journal = {IEEE Transactions on Visualization and Computer Graphics (Proc. IEEE VIS)},
	author = {Yi, Ji Soo and Kang, Youn ah and Stasko, John and Jacko, J.A.},
	year = {2007},
	note = {Conference Name: IEEE Transactions on Visualization and Computer Graphics},
	pages = {1224--1231},
}

@article{bostock_d3_2011,
	title = {D3: {Data}-{Driven} {Documents}},
	url = {http://vis.stanford.edu/papers/d3},
	doi = {10.1109/TVCG.2011.185},
	journal = {IEEE Trans. Visualization \& Comp. Graphics (Proc. InfoVis)},
	author = {Bostock, Michael and Ogievetsky, Vadim and Heer, Jeffrey},
	year = {2011},
}

@misc{w3c_wai_2019,
	title = {{WAI} {Web} {Accessibility} {Tutorials}: {Complex} {Images}},
	url = {https://www.w3.org/WAI/tutorials/images/complex/},
	journal = {World Wide Web Consortium},
	author = {W3C},
	year = {2019},
}


\end{document}